# An Indirect Social Trust Model for Vehicular Social Networks Using Evolving Graph Theory

Max Hashem Eiza
School of Computer Science and Mathematics
Liverpool John Moores University
Liverpool, UK
M.HashemEiza@ljmu.ac.uk

Vinh Thong Ta
Department of Computer Science
Edge Hill University
Ormskirk, UK
tav@edgehill.ac.uk

*Abstract*—The increasing importance and consequent challenges of establishing indirect trusted relationships in highly dynamic social networks such as vehicular social networks (VSNs), are investigated in this paper. VSNs are mobile social networks that aim to create social links among travellers on the roads. Besides matching interests between two users, social trust is essential to successfully establish and nurture a social relationship. However, the unique characteristics of VSNs pose many challenges such as uncertainty, subjectivity and intransitivity to indirect social trust modelling. Furthermore, the current trust models in the literature inadequately address trust propagation in VSNs. We propose a novel indirect social trust model for VSNs using evolving graph theory and the Paillier cryptosystem. We consider the VSN as a highly dynamic social evolving graph where social ties among vehicles hold a trustworthiness factor that evolves over time. This factor is estimated based on the behaviours, opinions, distances, and communication metrics of the parties involved. Employing the homomorphic property of the Paillier cryptosystem, the proposed model targets the subjectivity problem when combining multiple opinions to establish an indirect trusted relationship. Through analysis of computational and communication complexities, we show the viability of the proposed model and the efficiency of its indirect trust computation algorithm.

*Keywords—Evolving Graph; Indirect Trust Model; Social Trust; Social Evolving Graph (SEG); Vehicular Social Network (VSN);*

## I. INTRODUCTION

Nowadays, social networks embody one of the most popular ways for people to communicate, where over 2.95 billion social network users are expected by 2020 [1]. Thanks to the modern and ubiquitous connected mobile devices, traditional online social networks (OSNs) such as Facebook and Twitter are not the only players in this phenomenon. Recently, Mobile Social Networks (MSNs) have become more popular as mobile devices take advantage of their close proximity and leverage different communication technologies such as Bluetooth and Wi-Fi Direct to socially interact. MSN represents an attractive option for supporting social interactions in a number of mobile environments, where it can take advantage of both infrastructure-based and opportunistic wireless networks [2].

More recently, as one of the main application domains of MSNs, Vehicular Social Networks (VSNs) have emerged as a new social interaction paradigm. VSNs are decentralised opportunistic communication networks formed among vehicles on the roads [3]. They represent a unique form of localised mobile social network that exploits vehicular communication links and offers travellers the opportunity to engage in social activities along the road. Direct inquiry of others with similar experience in proximity over social networks tends to be the most convenient and efficient approach to acquiring up-to-date, specialised and domain-specific information for travellers [4]. Moreover, a TripAdvisor survey shows that 76% of travellers share their travel experience including photos and clips via social networks and 52% do that while on the road [5].

Based on the physical and social distances of users, many applications have been proposed in the context of VSNs [6] (e.g., CarPlay [7], UberPool [8], Waze [9], Verse [10], *etc.*). While some of the aforementioned applications require an Internet connection (i.e., infrastructure-based), the focus of this paper is on ad-hoc VSNs (i.e., infrastructure-less). Ad-hoc VSNs are constructed in a distributed fashion where each vehicle is responsible for maintaining its social relationships with other vehicles on the road. This approach offers high flexibility, especially on highways and requires no infrastructure support. However, the trustworthiness of information shared with each other is important, particularly, when the correctness of the information is critical, such as traffic information (e.g., Waze), or ride-sharing information (e.g., Uberpool).

In VSNs, the issue of trust is even more crucial as communication links between vehicles are usually short-term, and a reliable trust management approach is indispensable to measure the trustworthiness of the received information. While establishing a direct trusted relationship between two vehicles can be achieved based on their similarities and direct interactions, indirect trust computation represents the main challenge for social trust modelling in VSNs. In the literature, indirect trust between two vehicles is computed based on the opinions of other vehicles (i.e., using the propagative nature of the trust). Jsang [11] suggested using the subjective logic model to accommodate uncertainty and belief while computing indirect trust. Hence, to reflect the uncertainty and subjectivity of users' assessment of each other's trustworthiness, an opinion in subjective logic has three components: belief, distrust and uncertainty [6].

Nonetheless, this approach faces many challenges when handling complex topologies (i.e., large global social networks) in terms of trust computation efficiency. In addition, nodes' opinions can be easily influenced when observing others' opinions (i.e., the problems of information cascading and oversampling [12]). To rectify these problems, we propose a novel indirect trust model for VSNs using evolving graph theory. Employing the evolving graph theory in synergy with the Paillier cryptosystem, this paper makes two novel contributions. First, the proposed social evolving graph (*SEG*) reduces significantly the complexity of modelling decentralised trust in VSNs. A VSN is considered a highly dynamic graph where nodes represent vehicles and time-varying edges indicate connections and/or social relationships between vehicles. Secondly, using the homomorphic property of the Paillier cryptosystem, the problems of information cascading and oversampling are addressed when computing indirect trust based on other vehicles' opinions.

The rest of this paper is structured as follows. Section II states the preliminaries employed in this paper. Section III describes the social evolving graph (*SEG*) model for VSNs. Section IV introduces the indirect trust model using the proposed *SEG* model for VSNs and Paillier cryptosystem. Section V provides a computation and communication complexities analysis of the proposed model. Section VI highlights the benefits of the proposed model through a brief comparison with related work. Finally, Section VII concludes the paper and discusses future work.

## II. PRELIMINARIES

### A. Social Trust and Basics of Social Theory

Social trust can be defined in terms of two aspects of a relationship between two parties: risk and interdependence [13]. The risk represents the uncertainty about the intentions of the other party, while interdependence means that the interests of the two parties are intertwined and neither of them can achieve its own interests without the other. Any change in one of these aspects would alter the established relationship and may make it either stronger or weaker.

Since a social network can be modelled as a directed communication graph, social theory defines several aspects that can be utilised to localise the most significant nodes and quantify their relative importance to other nodes [14, 15]. Understanding these aspects is essential to grasp the foundations of social interactions in social networks. In the following, we focus on three of these aspects that directly affect the node trust in VSNs. The full set can be found in [16].

*1) Propinquity*: Under equal conditions, propinquity means that if two nodes are geographically close, they are more likely to be socially connected.

*2) Homophyly*: homophyly is the common social attributes (i.e., the similarity) between two nodes such as having the same favourites, having the same travelling destination, *etc*. Thus, it is more likely that nodes (i.e., travellers) with the same social attributes to have a connection [15]. Let $HP_i = \{S_{i,x}\}$ where $x = 1…w$ and $S_{i,x} \in \{0,1\}$ be the social interests profile of $v_i$. The homophyly between $v_i$ and $v_j$ can be evaluated as follows [10]

$$SHP_{ij} = \frac{\sum_{k=1}^{w} S_{i,k} S_{j,k}}{\sqrt{\sum_{q=1}^{w} S_{i,q}^2} \cdot \sqrt{\sum_{y=1}^{w} S_{j,y}^2}} \quad (1)$$

For instance, based on the following social interests {Football, Rap music, Mountain climbing} and $HP_i = \{1, 1, 0\}$, we can say that $v_i$ likes watching football, and listening to rap music but does not like mountain climbing. If $SHP_{ij} = 1$, then $v_i$ and $v_j$ have the same social interests, thus they are more likely to socially communicate. Otherwise, if $SHP_{ij} = 0$, then $v_i$ and $v_j$ have no interests in common and it is unlikely for them to have a social connection. The evaluation of the homophyly $SHP_{ij}$ factor between two nodes is the first step to determining the interdependence aspect of any possible social interaction between them.

*3) Degree Centrality:* A central node in a communication graph is the one that has a large number of in-links and out-links (i.e., connections) with other nodes. The degree of a node $v_i$, denoted as $C_D(v_i)$, can be calculated as follows [15]:

$$C_D(v_i) = \frac{\sum_{\forall i \neq j} x_{ij}}{(W-1)} \quad (2)$$

where $x_{ij} = 1$ if $v_i$ is incident to $v_j$ and $x_{ij} = 0$ otherwise, and $W$ is the total number of nodes in the network. Nodes with a high degree of centrality project higher levels of trust for other nodes and encourage them to socially interact.

### B. Evolving Graph Theory

The evolving graph theory is proposed as a formal abstraction for dynamic networks [17]. The evolving graph is an indexed sequence of $\eta$ sub graphs of a given graph, where the sub graph at a given index corresponds to the network connectivity at the time interval indicated by the index number, as shown below in Fig. 1.

As depicted in Fig. 1, the edge {*A, B*} exists in the time interval [1–3], while {*A, C*} only exists at time interval 4. A journey in the evolving graph is the path in the underlying graph where its edges' time labels are in increasing order [18]. For instance, in Fig. 1, {*A, D, C*} is not a valid journey since edge {*D, C*} exists only in the past with respect to edge {*A, D*}. Further, it is easy to see that {*A, B, E, G*} and {*D, C, E, G*} are valid journeys while {*D, C, E, G, F*} is not. Labelling edges with presence times makes the evolving graph suitable to model the dynamic nature of the vehicles' connections in VSNs.

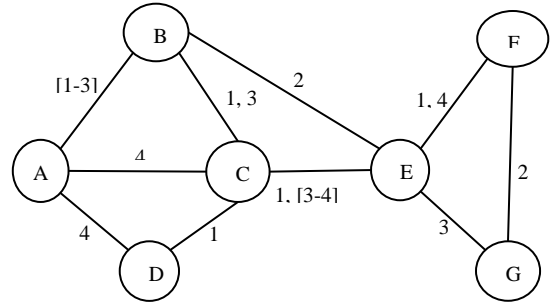

Fig. 1. Basic Evolving Graph Model [19]

### C. The Paillier Cryptosystem

The Paillier cryptosystem is a modular, public key encryption scheme with an additive homographic property [20]. Let $n = pq$ where $p$ and $q$ are large primes and $\lambda = lcm(p – 1, q – 1)$. Select $g \in \mathbb{Z}_{n^2}^*$ and calculate $\mu = (L(g^\lambda \bmod n^2))^{-1} \bmod n$ where $L$ is defined as $L(u) = (u – 1)/n$. The public key is $(n, g)$ while the private key is $(\lambda, \mu)$. The encryption process is carried out as follows. Let $m \in \mathbb{Z}_n$ be a message to be encrypted. Choose a random number $r \in \mathbb{Z}_n^*$ and compute $E(m, r) = c = g^m \cdot r^n \bmod n^2$ where $c$ is the ciphertext. To decrypt, compute $D(E(m, r)) = m = L(c^\lambda \bmod n^2) \cdot \mu \bmod n$.

Let $E(m_1, r_1)$ and $E(m_2, r_2)$ be the ciphertexts of messages $m_1$ and $m_2$, respectively, using the same public key $(n, g)$. In the Paillier cryptosystem, multiplying two ciphertexts will decrypt to the sum of their original plaintexts. Hence, $D(E(m_1, r_1) \cdot E(m_2, r_2)) \bmod n^2) = m_1 + m_2 \bmod n$.

## III. SOCIAL EVOLVING GRAPH (SEG) MODEL FOR VSNs

### A. Motivation

The current evolving graph theory cannot be directly applied to VSNs as the social trust evolution among connected vehicles cannot be known or scheduled in advance as time intervals. Also, the current evolving graph model does not consider the social theory aspects of communicating vehicles.

To facilitate the establishment of direct and indirect trusted social relationships among vehicles, we extend the current evolving graph model to develop a social evolving graph (*SEG*) model for VSNs. The *SEG* model aims to capture the social characteristics of the existing nodes and establish social connections among these nodes. Moreover, *SEG* facilitates efficient computation of the trustworthiness of social links between two vehicles that are not in close proximity (i.e., indirect trust computation). In the following, we introduce the proposed *SEG* model and explain its features in detail.

### B. SEG Model

In VSNs, establishing a new social connection between two vehicles does not only depend on being in close proximity (i.e., within the transmission range of each other). It also depends on their social attributes and interests. Thus, in the *SEG* model, each link is characterised with a set of attributes that describe it in terms of connectivity, social aspects and trustworthiness. The social link $SL_{ij}$ is only established between two vehicles $v_i$ and $v_j$ if it satisfies the following conditions. First, the $SHP_{ij}$ between $v_i$ and $v_j$ (i.e., the homophyly) should be higher than a predefined threshold $\Psi_H$, thus they share the minimum level of interest. Secondly, the expected communication link duration $ET_{ij}$ should be higher than a predefined time threshold $\Psi_L$ to ensure a meaningful social interaction. Finally, the trustworthiness of the potential social connection should satisfy the trust $\Psi_T$ threshold specified by involved parties as explained later in Section IV.

The value of $\Psi_H$ can be advertised by the vehicle itself along its *HP* profile (e.g., high value of $\Psi_H$ indicates that the user is only interested in communicating with other users that have a lot in common with himself/herself). On the other hand, the time threshold $\Psi_L$ can be determined by the current application. For instance, to share a video clip with other vehicles, the connection time should be long enough to watch/download the video file.

Fig. 2 shows an example of the *SEG* model on a highway at two-time instants $t = 0\ s$ and $t = 10s$ where $\Psi_H = 0.6$, $\Psi_L = 12s$ and $\Psi_T = 0.5$. Each node in Fig. 2 represents a vehicle on the highway. It can be seen in Fig. 2 that the proposed *SEG* model extends the traditional evolving graph model, in Fig. 1, by adding the tuple $(t, ET_{ij}, SHP_{ij}, TST_{ij})$ on edges, where $t$ denotes the current time, $ET_{ij}$ denotes the expected social link $SL_{ij}$ duration, $SHP_{ij}$ denotes the homophyly between $v_i$ and $v_j$, and $TST_{ij}$ denotes the trustworthiness factor of the potential social link between $v_i$ and $v_j$. Note that $TST_{ij} \neq TST_{ji}$ (i.e., trust is asymmetric since $A$ could trust $B$ more than $B$ could trust $A$). The asymmetric nature of direct trust is further illustrated in (4) when $TST_{ij}$ is estimated. In Fig. 2, communication links represent weak relationships (i.e., vehicles only communicate but no social links are established).

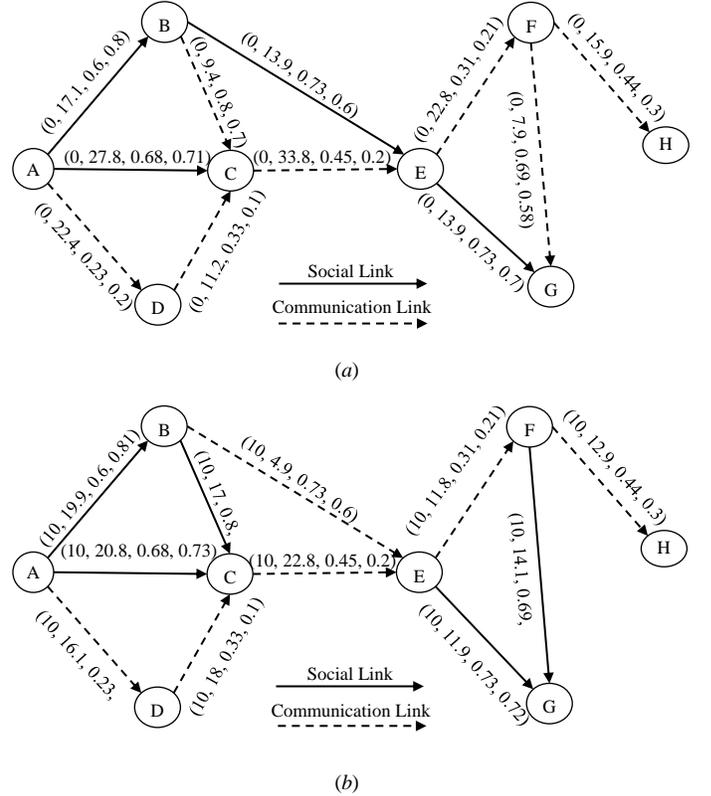

Fig. 2. Social Evolving Graph (*SEG*) Model at (a) $t = 0s$ and (b) $t = 10s$

The expected social link duration $ET_{ij}(t)$ between two nodes at time $t$ can be estimated as follows [21]:

$$ET_{ij}(t) = \frac{H - \theta\sqrt{(y_i(t) - y_j(t))^2 + (x_i(t) - x_j(t))^2}}{|v_i(t) - \vartheta v_j(t)|} \quad (3)$$

where $H$ is the communication range of each node, $\theta = -1$ and $\vartheta = 1$ when $v_j$ overtakes $v_i$, $\theta = 1$ and $\vartheta = 1$ when $v_i$ moves forward in front of $v_j$, $\theta = -1$ and $\vartheta = -1$ when $v_i$ and $v_j$ are moving toward each other, and $\theta = 1$ and $\vartheta = -1$ when $v_i$ and $v_j$ are moving away from each other.

In order to estimate the trustworthiness factor of the social link $TST_{ij}$, we modify the subjective logic approach suggested by Jsang [11] as follows. Instead of utilising the three components: belief, distrust and uncertainty, we propose to use two components only: belief and distrust. The rationale behind this modification is that most existing direct trust datasets only present the value of belief (e.g., the level of trust is 0.65). However, it is not clear whether the remaining 0.35 represents distrust or uncertainty or both [6]. Hence, we consider two components only (i.e., either trust or distrust where the latter includes uncertainty). Moreover, we integrate the degree

centrality $C_D(v_i)$ and homophyly $SHP_{ij}$ factor in estimating the $TST_{ij}$ as follows:

$$TST_{ij}(t) = (\delta_D C_D(v_i)) \cdot (\delta_H SHP_{ij}) + \delta_T TST_{ij}(t`) \quad (4)$$

where $0 \leq \delta_D \leq 1$, $0 \leq \delta_H \leq 1$ and $-1 \leq \delta_T \leq 0.1$ are weighting factors for $C_D(v_i)$, $SHP_{ij}$ and $TST_{ij}(t`)$, respectively, where $TST_{ij}(t`)$ is the trust value estimated at $t` < t$ based on previous interactions with $v_i$.

It is worth noting that $TST_{ij}(t)$ can be negative if $\delta_T < 0$. When $TST_{ij}(t) = 0$, it indicates complete distrust. However, when $TST_{ij}(t) < 0$, it indicates that the social interaction experience $v_i$ had with $v_j$ in the past is not pleasant. Thus, we allow negative trust values to discourage other nodes to socially interact with this node or consider its opinion credible. Thus, encouraging all vehicles to adopt a positive behaviour when interacting with others. It is worth noting that $TST_{ij}(t`)$ may not be available especially when vehicles use different pseudonymous identities on the roads. However, depending on the depth of the social relationship between two vehicles, they might reveal their real identities to each other. This issue requires more investigation and is left for future work.

The reason behind integrating both degree centrality $C_D(v_i)$ and homophyly $SHP_{ij}$ in estimating $TST_{ij}$ in (4) is related to the people's perception of social similarities and social connections with others. For instance, the fact that two travellers support the same football team would have an impact on their perception of each other and, consequently, the social trust level between them. Besides, knowing that the communicating node has a large number of social connections with others contributes to the feeling that this node can be trusted.

In the $SEG$ model, the social link $SL_{ij}$ between two vehicles is available if and only if $SHP_{ij} > \Psi_H$ and $ET_{ij} > \Psi_L$ and $TST_{ij} > \Psi_T$. For instance, $\{A, D\}$ in Fig.2 (a) is not a social link because it does not satisfy the condition of the homophyly as $0.23 < 0.6$. Let $Estab(SL_{ij})$ be a function that determines whether $SL_{ij}$ can be established or not. Thus, we can write:

$$Estab(SL_{ij}) = \begin{cases} True & if \ SHP_{ij} > \Psi_H \wedge ET_{ij} > \Psi_L \wedge TST_{ij} > \Psi_T \\ False & otherwise \end{cases} \quad (5)$$

In Fig. 2(a), It can be noticed that the following social links are established $\{A, B\}$, $\{A, C\}$, $\{B, E\}$ and $\{E, G\}$. After 10 seconds, in Fig. 2 (b), the set of the established social links changes to $\{A, B\}$, $\{A, C\}$, $\{B, C\}$, $\{E, G\}$ and $\{F, G\}$. It is worth noting that the trustworthiness of the following social links $\{A, B\}$, $\{A, C\}$ and $\{E, G\}$ has evolved over time from $t = 0s$ to $t = 10s$ (i.e., the social relationship is nurtured).

In VSNs, we assume that each vehicle along the road constructs its own version of the $SEG$ model shown in Fig. 2. This is possible when vehicles utilise the information received within the basic safety messages (BSMs) that are periodically exchanged in vehicular networks. BSMs are essential parts of the 5.9-$GHz$ dedicated short-range communication (DSRC) vehicular communication standard deployed [22]. Hence, each vehicle $v_i$ can be only concerned with nodes of interest (i.e., nodes that share the same social attributes/interests).

## IV. SEG-BASED INDIRECT SOCIAL TRUST MODEL FOR VSNS

In the following, we propose an intuitive indirect trust model based on the $SEG$ model to compute the trustworthiness of an indirect social link between two nodes that are not in close proximity. The indirect trust computation relies on the trustworthiness of social links among parties involved and their credibility based on their neighbours' opinions.

### A. SEG- Dijkstra's Algorithm

Let $s$ and $d$ be two vehicles that want to establish a social connection but are not in close proximity. In order to do so, the trustworthiness of this social link must be estimated. Prior to that, $s$ needs to find all possible routes to $d$. Finding all possible routes between two nodes in a social communication graph is an NP-hard problem. Therefore, we propose $SEG$-$Dijkstra$ a modified version of the Dijkstra algorithm [23] to utilise in the $SEG$ model in Section III.

Based on the $SEG$ model at $v_i$, $SEG$-$Dijkstra$'s algorithm eliminates all links that do not satisfy the predefined time threshold $\Psi_L$. After that, it finds the set of routes $M_{sd} = \{R_1, R_2 \ldots R_z\}$ that connect $s$ to $d$ in $SEG$. In the following, pseudocode for the $SEG$-$Dijkstra$ algorithm is provided.

> **Input:** An SEG Model and a source vehicle s.
> **Output:** Array $\Re$ that contains the routes from s to d.
> **Variables:** A set Q of unvisited vehicles.
> **1.** Eliminates all links where $ET < \Psi_L$;
> **2.** Initialize array Q by inserting s;
> **3.** While Q is not empty do
> (a) $x \leftarrow$ the node with the longest ET in Q;
> (b) Mark x as visited;
> (c) For each open neighbour y of x do
>   1. Set $\Re[y] \leftarrow x$;
>   2. Insert y if not visited in Q;
> (e) Close x;
> **4.** Return the array $\Re$;

### B. Indirect Trust Computation Algorithm

Since $s$ calculates $M_{sd}$, the indirect trust computation algorithm starts. First, $s$ sends an opinion request to each one-hop neighbour $v_d$ of $d$ (i.e., nodes that have a direct social relationship with $d$). The request contains: 1) the list of nodes from $s$ to the neighbourhood where each step is weighted according to its distance from $s$. These weights contribute to formulating the final opinion collected from these nodes as explained later; 2) an opinion value $E(OP_s, r_s)$ that is initialised and encrypted by $s$ using the Paillier cryptosystem with $OP_s = 0$.

Upon receipt of the request, $v_d$ formulates its opinion $OP_{vd}$ as follows: $OP_{vd} = \delta_{OPv}TST_{vd}$ where $0 < \delta_{OPv} \leq 1$ is the weight factor associated with node $v_d$ and $TST_{vd}$ is the trustworthiness of $v_d$'s social link with $d$. Before using the Paillier cryptosystem, $v_d$ should convert its opinion to an integer (e.g., 0.58 is converted to 58). After that, it adds its opinion as $E(OP_{vd}, r_{vd}) \cdot E(OP_s, r_s)$ (i.e., two ciphertexts multiplication). Subsequently, it sends the request backwards one step to $s$ (i.e., to the node that is two hops away from $d$). The two-hop away node will perform the same operation and sends the request backwards until $s$ is reached. This way, each node's opinion will not be affected by other nodes' opinions because it is hidden (i.e., encrypted). Finally, $s$ decrypts the final aggregated opinion $OP_F$, which is according to the homomorphic property, the sum of all collected opinions. Subsequently, $s$ calculates $TST_{sd}$ as follows

$$TST_{sd}(t) = (\delta_D C_D(d)) \cdot (\delta_H SHP_{sd}) + \delta_F \left( \frac{OP_F}{|R_{sd}|} \right) \quad (6)$$

where $0 \leq \delta_H \leq 1$ is the weight factor for opinions of nodes along the route $R_{sd}$ from $s$ to $d$. Note that $s$ has to revert the final aggregate opinion from integer while applying (6) because $TST_{sd} \leq 1$ (e.g., 52 becomes 0.52).

It is worth noting that the weights $\delta_{OPv}$ associated with each step starts with 1 and decreases with each step away from $s$. The reason behind this is related to the chain of trust of nodes' opinions from $s$. Let $R_1 = \{s, A, E, d\}$ and $R_2 = \{s, C, G, d\}$ be two routes between $s$ and $d$ in Fig. 3 where $s$ and $d$ wish to establish a social connection. Each social link is associated with the direct trust value where it can be noticed that $s$ trusts $A$ more than it trusts $C$. Conceptually, $A$'s opinion is trusted by $s$ more than $C$'s opinion. Therefore, the opinion from a direct connection is given the maximum weight. The far we move from $s$, the less we trust nodes' opinions. Although $G$ and $E$ trust $d$ the same, $C$ trusts $G$ with 0.9 while $s$ trusts $C$ with 0.4. Thus, $C$'s opinion about $G$ is trusted less by $s$. the opposite argument applies to $A$'s opinion about $E$.

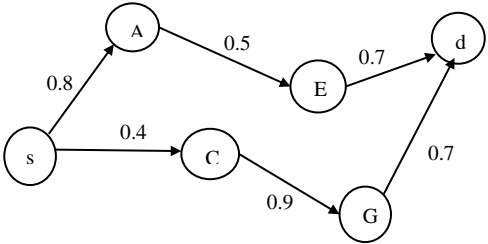

Fig. 3. Indirect trust computation algorithm – chain of trusted opinions

## V. ANALYSIS OF COMPUTATIONAL AND COMMUNICATION COMPLEXITIES

In our proposed model, the computation complexity incurs in two components: 1) the *SEG-Dijkstra's* algorithm that finds all possible routes between $s$ and $d$; and 2) the indirect trust computation between $s$ and $d$.

Let the number of nodes be $|V|$ and the number of edges be $|E|$ in the *SEG*. The while loop at *step* 3 in *SEG-Dijkstra's* algorithm is executed $|V|$ times. In *step* 3(a), we extract the node with the longest *ET* in $Q$ hence each node will be added to $Q$ once and deleted from $Q$ once. In the worst case, this step takes $O(|V|)$ for all nodes. However, if $Q$ is implemented as a heap, then the computational complexity can be reduced to $O(log|V|)$. The edge relaxation process and updating the array $\Re$ in *steps* 3(c)–3(e) takes $O(|E|+|V|)$. Since the *SEG-Dijkstra's* algorithm is proposed to work in the vehicular social communication graph on highways (i.e., *SEG* is a sparse graph), we can conclude that the total computational complexity of *SEG-Dijstra's* algorithm is $O((|E|+|V|)log|V|)$.

However, if more vehicles enter the highway and more social links are established among vehicles, (i.e., the sparseness of *SEG* decreases), the computational complexity becomes $O(|V|^2 log|V|)$. However, the number of vehicles that can enter the highway is controlled by the highway capacity. Note that *SEG-Dijstra's* algorithm is applied on the *SEG* model at the source node, thus there are no messages needed (i.e., no communications complexity).

Finally, we discuss the complexity of the *SEG*-based indirect trust computation algorithm in terms of communication and computational complexities. In the context of communication complexity, the worst case complexity is $O(|V| + |E|)$ messages (i.e., requests for nodes' opinions from the source node $s$). Traversing the longest route in the network, which contains at most $|V|$ nodes and $|V-1|$ links, requires $O(V)$ messages. Therefore, it requires $O(|V| + |E|)$ messages to receive opinions about all possible routes from $s$ to $d$.

In terms of the computational complexity, upon receiving an opinion request, each node performs one encryption and one multiplication. These two operations are carried out by each node. In Paillier cryptosystem, the encryption requires two exponentiations mod $n^2$ (without pre-computation). On the other hand, decryption, which is only carried out by $s$ when it receives the final opinion, essentially requires one exponentiation mod $n^2$. Note that $m$ is integer and, in this context, $m$ is not a large integer. On Intel Pentium 4 3.0-GHz machine, with pre-computation, the time needed to perform a single exponentiation operation is $T_{exp} = 1.1$ *ms* [24]. Thus, utilising Paillier cryptosystem will not affect the efficiency of the proposed indirect trust model.

## VI. COMPARISON WITH RELATED WORKS

Trust modelling and management in Vehicular Ad-hoc Networks (VANETs) and VSNs are new and intensively researched areas [25–30]. However, most of the studies focused on centralised or infrastructure-based solutions, while only few works have addressed distributed trust management dealing with direct trust issue only. To the best of our knowledge, an efficient indirect trust model has not been proposed yet for VSNs. Our work differs from the studies below as it proposes an efficient and distributed indirect social trust model for VSNs.

Liao *et al.* [30] proposed a secure incident reports approach in VANETs to increase safety on roads. The likelihood of the accuracy of Vehicle-to-Vehicle incident reports is estimated based on the trustworthiness of the report originator and the forwarding vehicles. Information about the vehicle's behaviour is collected in a crowd-sourced fashion from the road-side units, managed by central traffic authorities, using Vehicle-to-Infrastructure communications. The trust scores of the reports are computed by a central authority by aggregating the vehicle's behaviour history based on incident report accuracy. In [6], the authors outlined a distributed solution for trust management in VSNs. The proposed solution suggests that each vehicle maintains a local social connection tree (SCT), which will be updated according to its social connection history. The root of a local SCT is the vehicle that maintains this SCT, while the edges are the social connections of this vehicle. Vehicles can build and refine their global social networks based on the SCTs received from others. This approach focuses only on direct trust, without giving any details about the trust calculation.

A trust propagation model/augmented trust model for vehicular networks is proposed in [27]. Each vehicle is assigned a single, long-term trust value instead of a short-term changing value. They store the trust value of each vehicle centrally and use PKI-based trust propagation to update the trust values. The authors in [28] proposed a reputation mechanism based on user behaviors and some historic features. The reputation of each vehicle is based on their message communication, and the current reputation is estimated based on previous interactions/reputation. Finally, in [29] a trust evaluation algorithm for VSNs and a three-level cloud-based VSN architecture applying this trust algorithm have been proposed, with a performance analysis carried out with PEPA (Performance Evaluation Process Algebra).

VII. CONCLUSION & FUTURE WORK

In this paper, the importance of social indirect trust in establishing indirect social connections in VSNs has been investigated and the impact of the highly dynamic nature of VSNs identified. To establish reliable and trusted social relationships among travellers in VSNs, a novel indirect social trust model has been proposed using the evolving graph theory in synergy with the Paillier cryptosystem. The proposed model synergistically exploits the evolving graph model properties to build dynamic social evolving graphs. Moreover, it achieves a high level of objectivity when estimating indirect trust between two users by using the homomorphic additive feature of the Paillier cryptosystem. It was shown that the proposed model has low communication and computation complexities when applied to VSNs in a highway environment. For future work, the intention is to implement the proposed model with added features such as context classification where vehicles can maintain different *SEGs* for different applications. Also, we intend to perform extensive simulation experiments using real-life trust datasets.